\documentclass[a4paper]{article}
\usepackage{amsmath}
\usepackage{amssymb}

\begin{document}

\title{Nonclassical symmetry and Riemann invariants}
\author{Souichi Murata\footnote{Corresponding author.E-mail address: smurata@r.phys.nagoya-u.ac.jp} \\
Department of Physics, Nagoya University, Chikusa-ku, Nagoya, 464-8602, Japan
}

\date{}

\maketitle
\begin{abstract}
In this paper it is shown that Riemann invariants are invariant under nonclassical symmetries of a hyperbolic system. As a specific example, we study the one-dimensional shallow water equations on the flat and present another type invariance under nonclassical symmetries.\\\\
Keywords:Riemann invariant; Nonclassical symmetry; Shallow water equations
\end{abstract}

\section{Introduction}

\ \ \ \  The method of Riemann invariants is based on the method of characteristics \cite{Courant,Taniuchi,Grandland}. The main feature of this method is that we can introduced new dependent variables (called Riemann invariants) which remain constant along appropriate characteristic curves of the hyperbolic system of partial differential equations. This method enables us to reduce the number of dependent variables, so that procedure of solving the system is simplified. 
But,  Riemann invariants do not exist in the hyperbolic system having more than three dependent variables \cite{Taniuchi} .

The notion of the nonclassical symmetry was introduced by Bluman and Cole to extend  the classical method for symmetry reduction \cite{Baumann}. In  the nonclassical symmetry analysis the original PDEs are augmented with the invariant surface conditions associated to  infinitesimal generators. By requiring that the original PDEs and the invariant surface conditions are invariant under infinitesimal transformations yields an overdetermined, nonlinear system of equations for the coefficients in the infinitesimal generators. These equations for the coefficients are called the determining equations.Since the number of determining equations arising in the nonclassical symmetry analysis  is smaller than for the classical method and since all solutions of the classical determining equations necessarily satisfy the nonclassical determining equations, the nonclassical symmetry may be larger in the classical case\cite{Olver, Bluman, CRC}.

In this paper, we will seek another type of invariance that substitutes for Riemann invariants. First of all, we  show that Riemann invariants are one of  invariance under nonclassical symmetries of the hyperbolic system. As a specific example, we consider nonclassical symmetries of the shallow water equations \cite{CRC, Lou, Stoker}. These equations describing an incompressible heavy fluid flow with a free surface on flat plate coincide with those of a polytropic gas with the specific heat ratio equals to two. In this example, we will present an another type of invariance under the nonclassical symmetries.

\section{Mathematical description of symmetry analysis}

\ \ \ \ First of all, we illustrate classical symmetry analysis.

A system of n-th order PDEs in $p$ independent and $q$ dependent variables is given as a system of equations

\begin{align}
\Delta_{\nu}(\mathbf{x},\mathbf{u}^{(n)})=0, \quad \nu= 1 \dots l, 
                                                            \label{PDEs}
\end{align}  
involving $\mathbf{x}=(x^1, \dots, x^p)$, $\mathbf{u}=(u^1,\dots,u^q)$ and the derivatives of $\mathbf{u}$ with respect to $\mathbf{x}$ up to n, where $\mathbf{u}^{(n)}$ represents all the derivatives of $\mathbf{u}$ of all orders from 0 to n.  

We consider a one-parameter Lie group of infinitesimal transformations acting on the independent and dependent variables of the system \cite{Olver, Bluman}

\begin{align}
\tilde{x}^i &=x^i+\epsilon \xi^i(\mathbf{x},\mathbf{u})+O(\epsilon^2),
                                                       \quad i=1,\dots,p, \\
\tilde{u}^j &=u^j+\epsilon \Phi^j(\mathbf{x},\mathbf{u})+O(\epsilon^2),
                                                       \quad j=1,\dots,q. 
\end{align}
The infinitesimal generator $\mathbf{V}$ associated with the above group of  transformations can be written as 

\begin{align}
\mathbf{V}=\sum^{p}_i \xi^i(\mathbf{x},\mathbf{u})\partial_{x^i}
                +\sum^{q}_j \Phi^j(\mathbf{x},\mathbf{u})\partial_{u^j}.
                                                        \label{generator} 
\end{align}
The Invariance of the system (\ref{PDEs}) under the infinitesimal transformations leads to the invariance condition

\begin{align}
Pr^{(n)}\mathbf{V}[\Delta_{\nu}(\mathbf{x},\mathbf{u}^{(n)})]=0,
                          \quad \nu=1,\dots l, \ \ \mbox{whenever} \ \  \Delta_{\mu}(\mathbf{x},\mathbf{u}^{(n)})=0, \label{prolongation}
\end{align}
where $Pr^{(n)}$ is the n-th order prolongation of the infinitesimal generator.

Since the coefficients of the infinitesimal generator do not include derivatives of $\mathbf{u}$,  we can separate (\ref{prolongation}) with respect to derivatives and solve the resulting overdeterminied system of linear homogeneous PDEs known as the determining equations.  Our purpose for calculating symmetries is to obtain similarity solutions which are invariant under the symmetries. Applying the criterion of invariance, we get the invariant surface conditions

\begin{align}
\mathbf{V}(u^i-\overline{u}^i)\bigg|_{u^i=\overline{u}^i}=0, \label{invariant}
\end{align}
where $\overline{u}^j$ is the function of $\mathbf{x}$.  In order to solve the conditions (\ref{invariant}), we have to integrate the Lie equation corresponding to $\mathbf{V}$ 

\begin{align}
 \frac{dx^1}{\xi^1}= \dots = \frac{dx^m}{\xi^m}
=\frac{du^1}{\Phi^1}=\dots = \frac{du^n}{\Phi^n}.  \label{Lie-eq}
\end{align}
Then, Eq (\ref{invariant}) and solutions of (\ref{Lie-eq}) yield an invariant solution under the classical symmetry.

An essential observation by Bluman and Cole was that an invariant solution
 dose not only satisfy the system of PDEs (\ref{PDEs}) but also the invariant surface conditions (\ref{invariant}). The nonclassical method applies the classical method of symmetry analysis to the extended system  \cite{Baumann}

\begin{equation}
\begin{cases}
\Delta_{\nu}(\mathbf{x},\mathbf{u}^{(n)})=0, \quad \nu= 1 \dots l, \\
\mathbf{V}(u^i-\overline{u}^i)\bigg|_{u^i=\overline{u}^i}=0.   
                                               \label{nonclassical-PDEs}
\end{cases}
\end{equation}
The invariant criterion reads in the original form as

\begin{equation}
\begin{cases}
Pr^{(n)}\mathbf{V}[\Delta_{\nu}(\mathbf{x},\mathbf{u}^{(n)})]=0,
        \quad \nu=1,\dots l, \ \ \mbox{whenever} \ \  
                           \Delta_{\mu}(\mathbf{x},\mathbf{u}^{(n)})=0, \\
Pr^{(1)}\mathbf{V}[\mathbf{V}(u^i-\overline{u}^i)]=0,
        \quad \nu=1,\dots l, \ \ \mbox{whenever} \ \ u^i=\overline{u}^i, 
                                               \label{nonclassical-pro}
\end{cases}
\end{equation}
The second equations of (\ref{nonclassical-pro}) imposes no additional conditions because these equations are  satisfied identically and the determining equations for nonclassical symmetry are nonlinear equations.

\section{Riemann invariants and nonclassical symmetries}

\ \ \ \ In the present section we will show that Riemann invrinats are one of special type invariance under nonclassical symmetries of a hyperbolic system.

Let the 1-dimensional quasilinear hyperbolic system for the column vector $\mathbf{U}$ with $n$ components $\mathbf{U}=(u_1, \cdots, u_n)$ be given by

\begin{align}
\frac{ \partial \mathbf{U}}{\partial t}+\mathbf{M}(\mathbf{U})
                \frac{ \partial \mathbf{U}}{\partial x}=\mathbf{0},
                                                    \label{section3-1}
\end{align}
where $\mathbf{M}$ is a $n \times n$ matrix, the elements of which are functions of $u_1, \cdots, u_n$, the $n$ eigenvalues of $\mathbf{M}$ denoted by 
$\lambda_1, \cdots \lambda_n$ are all real values and distinct, and 
the corresponding right eigenvalues are all linearly independent.

Let us consider nonclassical symmetries generated by the following form

\begin{align}
\mathbf{V}=\partial_t+\xi \partial_x.  \label{section3-2}
\end{align}
Substituting the invariant surface conditions associated with the infinitesimal generator (\ref{section3-2}) into the system (\ref{section3-1}), we have 

\begin{align}
\bigg( \mathbf{M}-\xi \mathbf{E} \bigg)
                \frac{ \partial \mathbf{U}}{\partial x}=0, \label{section3-3}
\end{align}
where $\mathbf{E}$ denotes an  identity matrix. It is evident that  the coefficient $\xi$ is an eigenvalue of the matrix $\mathbf{M}$.

The determining equations (\ref{nonclassical-pro}) corresponding to the system (\ref{section3-1}) take the form

\begin{align}
\frac{\partial \mathbf{U}}{\partial x}
               \mathbf{V}\bigg( \mathbf{M}-\xi \mathbf{E}\bigg) =0.
                                                        \label{section3-4}
\end{align}
Since the all elements of the matrix $\mathbf{M}-\xi \mathbf{E}$ are the functions dependent variables, the equations (\ref{section3-4}) are satisfied automatically, so  the infinitesimal generator (\ref{section3-2}) is a basis of the nonclassical symmetries of the system (\ref{section3-1}).  

 The families of the characteristic curves of the hyperbolic system  (\ref{section3-1}) are the same one of the invariant surface conditions \cite{Taniuchi}. Therefore, the Riemann invrinats are invariance under nonclassical symmetries of the hyperbolic system. 

In next section, we consider the shallow water equations and present another type of  invariance under the nonclassical symmetries.

\section{Nonclassical symmetry analysis for the shallow water equations}

\ \ \ \  The one-dimensional shallow water equations on a flat bottom are written in the following form

\begin{align}
h_t+u h_x+h u_x=0, \label{cont} \\
u_t+u u_x+  h_x=0, \label{motion}
\end{align}
where the time $t$, the position $x$, the depth of the water $h$, the velocity $u$ are normalized by $t_0$, $x_0$,  $h_0$\  and \  $u_0=x_0/t_0$, respectively; $g$ is a constant of the free-fall acceleration and $u_0=\sqrt{g h_0}$.

If we assume that the coefficient of $\partial_t$ of the infinitesimal generator of nonclassical symmetries  does not identically equal zero, then for the infinitesimal generator 

\begin{align}
\mathbf{V}=\partial_t+\xi(x,t,h,u)\partial_x+\phi(x,t,h,&u)\partial_h
                     +\psi(x,t,h,u)\partial_u, \label{V}
\intertext{the invariant surface conditions are }
h_t+\xi h_x-\phi&=0,  \label{invariant-h} \\
u_t+\xi u_x-\psi&=0.  \label{invariant-u}  
\end{align}

Substituting equations (\ref{invariant-h}) and (\ref{invariant-u}) into the equations (\ref{cont}) and (\ref{motion}) leads to 

\begin{align}
 ( \mathbf{M}-\xi\mathbf{E}) \left(  
   \begin{array}{c}
        h_x  \\
        u_x  \\
   \end{array}\right)  
=
-\left(  
   \begin{array}{c}
        \phi  \\
        \psi  \\
   \end{array}\right),
\intertext{where}
E = \left(  
   \begin{array}{cc}
        1  &    0 \\
        0  & \  1 \\
   \end{array}\right), \quad   
M = \left(  
   \begin{array}{cc}
        u  &    h \\
        1  & \  u \\
   \end{array}\right).   \nonumber
\end{align}

The determinant of the matrix $\mathbf{M}-\xi\mathbf{E}$ classifies the determining equations for the infinitesimal generator (\ref{V})  into two cases.

In the case  $\mbox{det}(\mathbf{M}-\xi\mathbf{E})= 0$ the function $\xi$ and $\psi$ satisfy the relation

\begin{align}
(\xi,\phi)=(u-k,k\psi),\quad  \mbox{where}\ \ k&=\pm\sqrt{h} \label{xi-phi},
\intertext{and the determining equations for nonclassical symmetries  are}
-h\psi_h+k\psi_u+\frac{3}{4}\psi&=0, \label{prolongation-1} \\
\psi_t+u\psi_x+\psi\psi_u+k\psi_x-k\psi\psi_h&=0. \label{prolongation-2}
\end{align}

 The equations (\ref{prolongation-1}) and (\ref{prolongation-2})  lead to the nonclassical symmetries

\begin{align}
\mathbf{V}_1=\partial_t+(u+ \sqrt{h})\partial_x,\quad 
\mathbf{V}_2=\partial_t+(u- \sqrt{h})\partial_x.
 \label{symmetry}
\end{align}

Solving the Lie equation corresponding to the infinitesimal generator $\mathbf{V}_1$, we get a invariant solution 

\begin{align} 
y=x-(u &+ \sqrt{h})t,\quad u=2\sqrt{H}+r,\quad  h=H(y). 
\end{align}

Similarity,  we get the another solution corresponding to the generator $\mathbf{V}_2$

\begin{align}
y=x-(u &- \sqrt{h})t,\quad u=-2\sqrt{H}+s,\quad  h=H(y) 
\end{align}
where $r$ and $s$ which are constant along the Lie equation are Riemann invariants.

For $\mbox{det}(\mathbf{M}-\xi\mathbf{E}) \neq 0$, the determining equations for nonclassical symmeties  take the complicated form

\begin{multline}
\phi_t+u\phi_x-h B \phi_h-A\phi_u-A(\xi_t+u\xi_x-h B \xi_h-A\xi_u)  \\
+ \psi A+\phi B +h\{\psi_x+A\psi_h+B\psi_u-B(\xi_x+A\xi_h+B\xi_u) \}=0
                                                 \label{prolongation-phi}
\end{multline}
\begin{multline}
\psi_t+u\psi_x-h B \psi_h-A\psi_u-B(\xi_t+u\xi_x-h B \xi_h-A\xi_u)  \\
+ \psi B +\{\phi_x+A\phi_h+B\phi_u-A(\xi_x+A\xi_h+B\xi_u) \}=0
                                                 \label{prolongation-psi}
\end{multline}
where
\begin{align}
A=\frac{ (\xi-u)\phi+h\psi }{  (\xi-u)^2-h},\quad 
  \quad B=\frac{\phi+(\xi-u)\psi }{ (\xi-u)^2-h}.
\end{align}

The problem is how to determine the functions $\xi$, $\phi$ and  $\psi $.  It is difficult to find a general solution of the equations (\ref{prolongation-phi}) and (\ref{prolongation-psi}).

We  restrict the functions to finding nonclassical symmetries for which $\xi=u$, $\phi=a\sqrt{h}\psi$\ and\ $\psi=\psi(t,u,h)$, which the constant $a=\pm1$.
Then, the equations (\ref{prolongation-phi}) and (\ref{prolongation-psi}) are reduced to the equation 

\begin{align}
D \psi +a\sqrt{h}(A\psi_h+B\psi_u)+\frac{3}{2}B\psi=0, \label{section4-29}
\intertext{
where $D \equiv \partial_t -hB \partial_h -A\partial_u$. Solving  the equation (\ref{section4-29}) yields 
}
\psi=-\frac{1}{\frac{3a t}{2\sqrt{h}}+f(h,u)}. \label{section4-30}
\end{align}
where $f(u,h)$ is an arbitrary function of $h$ and $u$.

Let us find an invariant solution under the nonclassical symmetry of the following form

\begin{align}
\mathbf{V}=\partial_t+u\partial_x+a\sqrt{h}\psi\partial_h+\psi\partial_u.
                                                      \label{section4-31}
\end{align}

Since we can not integrate the Lie equation associated with the infinitesimal generator (\ref{section4-31}) ,  we have to treat invariant surface conditions (\ref{invariant-h}) and (\ref{invariant-u}).

The derivatives $u_x$ and $h_x$  are found from the equations (\ref{cont}), (\ref{motion}), (\ref{invariant-h}) and (\ref{invariant-u}): 

\begin{align}
u_x&=-\frac{\phi}{h}, \qquad h_x=-\psi. \label{section4-32}
\intertext{Solving the equations (\ref{invariant-h}), (\ref{invariant-u})\ and \ (\ref{section4-32}) give the another type of invariant solution}
y=x-(u+a\sqrt{h})t,& \quad      u=2a\sqrt{h}+\alpha, \quad h=H(y) 
\intertext{where $\alpha$  an arbitrary constant and  $H(y)$ is the function given by }
    H(y)&=\int \frac{1}{\ \ f(y) \ \ }dy.
\end{align}

In this case, the variable $y$ remains constant along the following characteristics curve 

\begin{align}
\frac{dx}{dt}=u+a\sqrt{h}.
\end{align}

\section{summary}

\ \ \ \ We show that Riemann invariants are one of in variances under the nonclassical symmetries admitted by the hyperbolic system. The key point of the proof is that the characteristic curves for the Riemann invariants  are the  same equation of the Lie equation of the nonclassical symmetries.  In addition , we consider the shallow water equation and present another type of  invariance under the nonclassical symmetries of the shallow water equations. The determinant of the matrix $\mathbf{M}-\xi\mathbf{E}$ classifies the nonclassical symmetry analysis into two cases. The first case  is for Riemann invariants and  the latter case  is for tha another type invariance.

\end{document}